\documentclass{egpubl}
\usepackage{eurovis2022}

\EuroVisShort  

\usepackage[T1]{fontenc}
\usepackage{dfadobe}  

\usepackage{cite}  
\BibtexOrBiblatex
\electronicVersion
\PrintedOrElectronic
\ifpdf \usepackage[pdftex]{graphicx} \pdfcompresslevel=9
\else \usepackage[dvips]{graphicx} \fi

\usepackage{egweblnk}

\usepackage{microtype}                 
\PassOptionsToPackage{warn}{textcomp}  
\usepackage{textcomp}                  
\usepackage{mathptmx}                  
\usepackage{times}                     
\usepackage{cite}                      
\usepackage{tabu}                      
\usepackage{booktabs}                  

\usepackage{csquotes}
\usepackage{fontawesome}
\usepackage{pgfplots}
\usepackage{xspace}
\usepackage{enumitem}

\usepackage{caption}
\usepackage{subcaption}
\usepackage[normalem]{ulem}
\usepackage{xcolor}
\definecolor{mred}{rgb}{.80,.12,.30}
\definecolor{grey}{rgb}{0.5,0.5,0.5}
\definecolor{Purple}{rgb}{.75,0,.85}
\definecolor{BlueGreen}{rgb}{.05,.59,.73}
\definecolor{pistachio}{rgb}{0.58, 0.77, 0.45}

\newif\ifnotes
\notestrue

\let\origcite\cite

\renewcommand{\cite}[1]{\ifnotes\mbox{\origcite{#1}}\else \origcite{#1}\fi}

\newcommand{\iTaskExampleMidSentence}{``\textit{Plot sweatshirt sales for months in 2016. Observe an interesting pattern in the chart. Find another item and year for which sales follow a similar pattern,}''\xspace}

\title[Inferential Tasks as an Evaluation Technique for Visualization]%
      {Inferential Tasks as an Evaluation Technique for Visualization}

\author[Suh et al.]
{\parbox{\textwidth}{\centering 
            A.\ Suh$^{1}$\orcid{0000-0001-6513-8447}
        and A.\ Mosca$^{2}$\orcid{0000-0002-9008-5516}
        and S.\ Robinson$^{1,3}$\orcid{0000-0002-7514-4215}
        and Q.\ Pham$^{1}$\orcid{0000-0003-1579-4942}
        and D.\ Cashman$^{4}$\orcid{0000-0003-4853-5701}
        and A.\ Ottley$^{5}$\orcid{0000-0002-9485-276X}
        and R.\ Chang$^{1}$\orcid{0000-0002-6484-6430}
    }
    \\
{\parbox{\textwidth}{\centering 
        $^1$Tufts University, Medford, MA, USA \\
        $^2$Northeastern University, Boston, MA, USA \\
        $^3$EditShare, Watertown, MA, USA \\
        $^4$Novartis Pharmaceuticals, Cambridge, MA, USA \\
        $^5$Washington University in St. Louis, St. Louis, MO, USA 
    }
}
}

\begin{document}



\teaser{
\centering 
 \includegraphics[clip, trim=0 10.8cm 4.3cm 0, width=.85\textwidth]{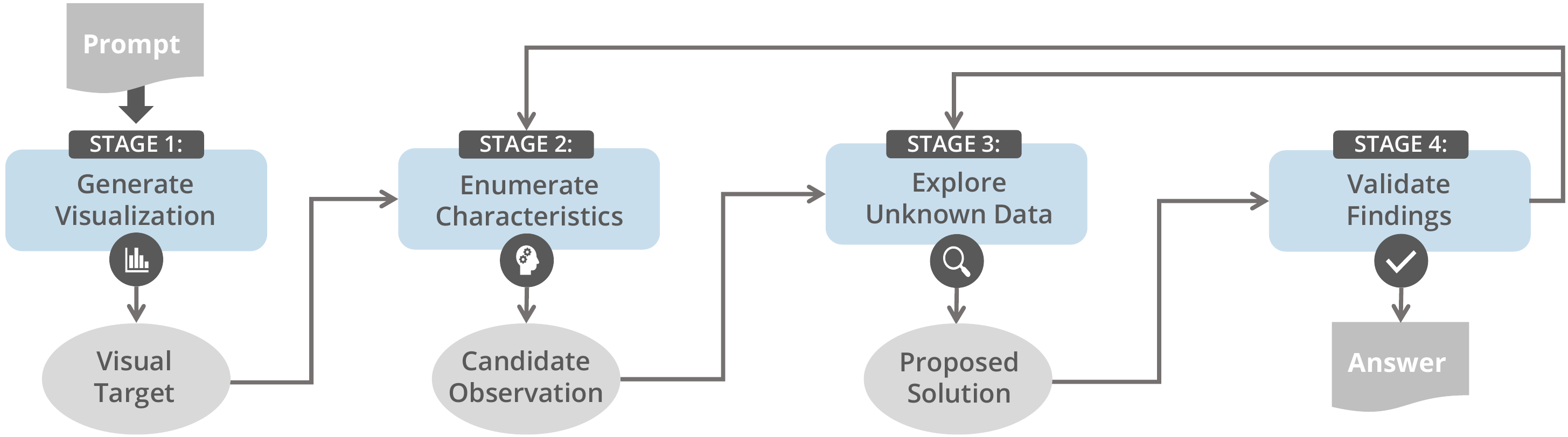} 
 \caption{\textit{The four distinct stages of evaluating interactive visualizations with an inferential task. Inferential tasks are based on the concept of \textit{inferential learning}, a process requiring users to rely on their problem-solving and reasoning abilities to draw conclusions that are not explicitly prompted.  In our proposed framework, users are given a prompt with the following instructions: build a specific visualization, then find a different subset of data that exhibits similar characteristics when visualized.  We describe each of these stages in Section~\ref{sec:itasks}.}
 }
 \label{fig:itask_teaser}
}

\maketitle
\begin{abstract}
Designing suitable tasks for visualization evaluation remains challenging. Traditional evaluation techniques commonly rely on `low-level' or `open-ended' tasks to assess the efficacy of a proposed visualization, however, nontrivial trade-offs exist between the two. Low-level tasks allow for robust quantitative evaluations, but are not indicative of the complex usage of a visualization. Open-ended tasks, while excellent for insight-based evaluations, are typically unstructured and require time-consuming interviews. Bridging this gap, we propose inferential tasks: a complementary task category based on inferential learning in psychology. Inferential tasks produce quantitative evaluation data in which users are prompted to form and validate their own findings with a visualization. We demonstrate the use of inferential tasks through a validation experiment on two well-known visualization tools.



\end{abstract}
\section{Introduction}
\label{sec:intro}

When evaluating interactive visualization systems, numerous approaches have been suggested over the years to ascertain both the benefits and limitations of a proposed tool\cite{chen2000empirical}.
Central to these approaches is the use of \textit{tasks} that users are asked to perform while using a visualization\cite{brehmer2013multi, dimara2018task}.  In 2005, Amar et al. presented ten low-level analytic tasks\cite{amar2005low} that remain commonly used in the evaluation of visualizations today\cite{isenberg2013systematic} (e.g., \cite{saket2018task, rosen2020linesmooth}). However, the use of low-level tasks in visualization evaluation poses a variety of challenges, such as the task's perceived lack of complexity, as well as the task's inability to capture a visualization's insight capabilities\cite{kosara2016empire, pandey2020towards}. 

North proposed the elimination of simple tasks altogether in experimental studies, instead suggesting ``complex benchmark tasks'' and insight-driven, ``open-ended protocols'' that more realistically assess the efficacy of visualizations\cite{north2006toward}. While open-ended tasks produce rich qualitative feedback\cite{peck2019personal, meyer2019criteria}, conducting and analyzing responses from think-aloud tasks through interviewing and open-coding is a time-consuming endeavor that is not always feasible or scalable for the designer\cite{anderson2010presenting, thudt2017expanding}. As a result, the ability to evaluate a visualization's insight capabilities both realistically and quantitatively has continued to be of interest to the research community\cite{battle2018evaluating, burns2020evaluate}. 

\begin{figure}[h!]
\centering 
 \includegraphics[clip, trim=0 5.5cm 9.6cm 0, width=.85\linewidth]{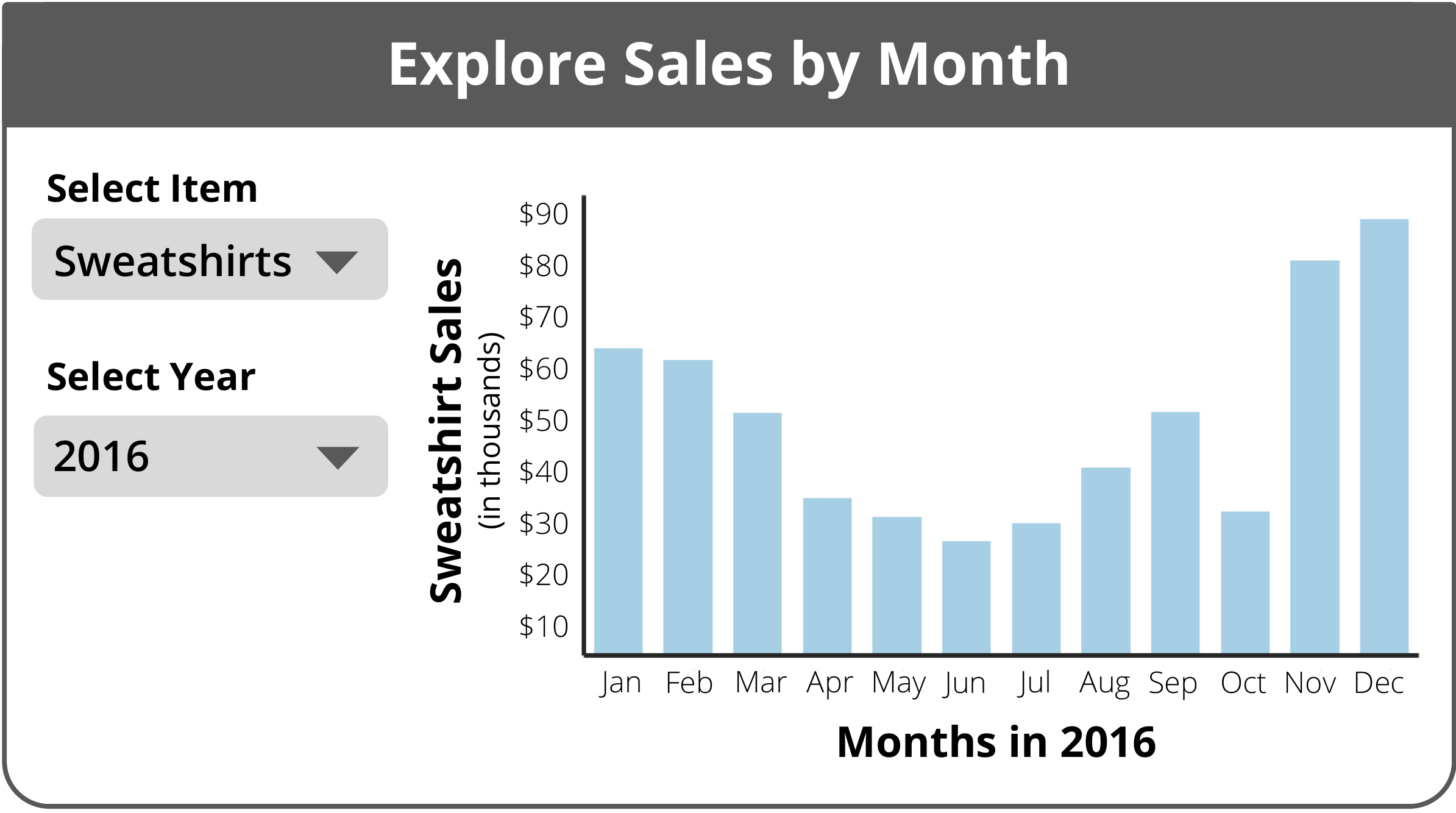} 
 \caption{\textit{Illustrative visualization tool that supports users in the exploration of monthly item sales for a specified year. This tool can be used to solve the inferential task: ``\textit{Plot sweatshirt sales by month for the year 2016. Observe how monthly sweatshirt sales are affected in 2016. Find another item, besides sweatshirts, whose monthly sales exhibit a similar relationship in the year 2016}.''}}
     \label{fig:inferentialTaskExample}
\end{figure}

In this paper, we formalize a new class of complex benchmark tasks that serve to complement open-ended protocols: \textit{Inferential tasks}. Inspired by the concept of inferential learning from psychology\cite{seel2012inferential}, inferential tasks require evaluation participants to construct knowledge by inferring relations between learned concepts and new observations. 
Moreover, inferential tasks can be set up with clear `correct' or `incorrect' answers, resulting in quantitative evaluation data that can be analyzed with the same statistical methods as low-level task evaluations. An example of an inferential task for visualization evaluation is shown in Figure~\ref{fig:inferentialTaskExample}.

Our motivation for proposing tasks requiring inferential learning is their success in prior visualization literature\cite{green2010using, ziemkiewicz2011locus}.
Green et al. and Ziemkiewicz et al. both showed that tasks that involve inferential learning produce more nuanced and informative evaluation data than tasks that only involve \textit{procedural learning} (i.e., traditional low-level tasks). Though the authors share these findings in their work, they do not offer guidelines for designing nor deploying inferential tasks in visualization evaluations.

We build on this previous work by defining a methodology for inferential tasks in visualization evaluation. We then demonstrate how our framework for inferential tasks can be used in practice in a validation experiment comparing two well-known exploratory visualization tools, Voyager 2\cite{2017-voyager2} and Polestar\cite{2016-voyager}. Our results indicate that the use of inferential tasks produce evaluation data that illuminates differences between the tools, while remaining straightforward to analyze quantitatively. Finally, we discuss design considerations, limitations, and future work. 

\begin{figure*} 
    \centering
    \subfloat[Many observations due to a vague prompt ]{{\includegraphics[width=0.3\textwidth]{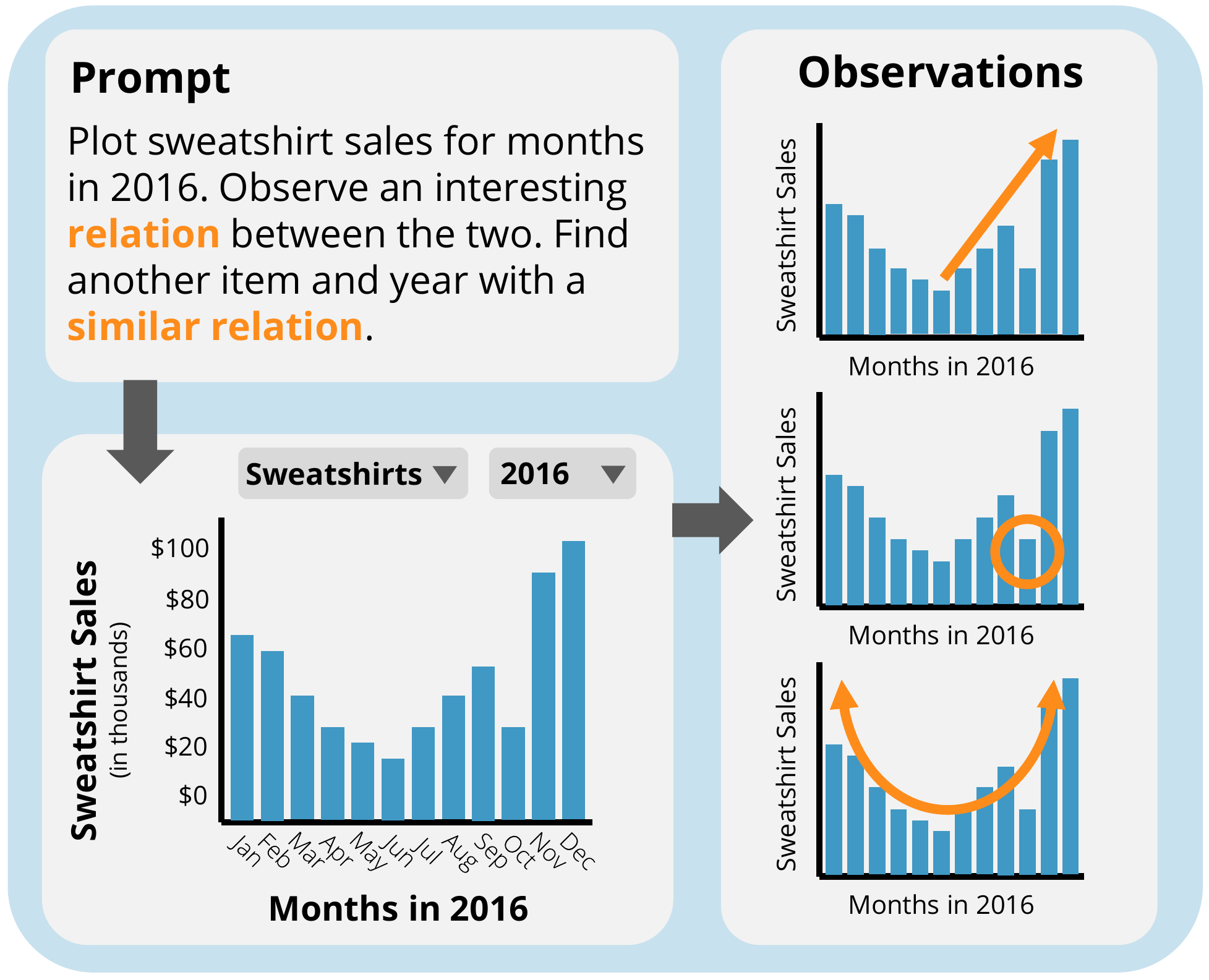} }
    \label{fig:tuning_hypothesis_space_a}
    }
    \quad
    \subfloat[Fewer observations due to a less vague prompt ]{{\includegraphics[width=0.3\textwidth]{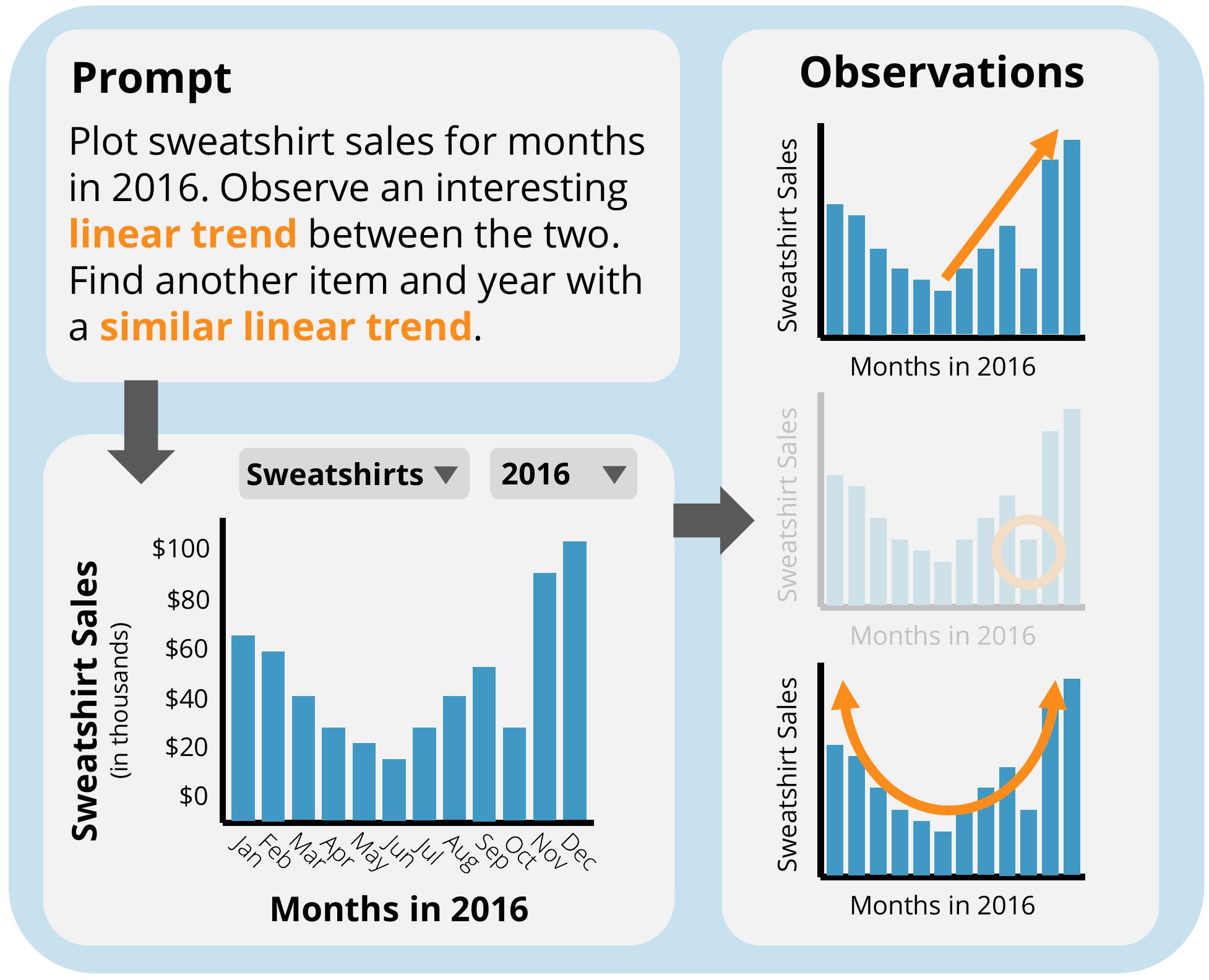} }
    \label{fig:tuning_hypothesis_space_b}
    }
    \quad
    \subfloat[Least observations due to a specific prompt]{{\includegraphics[width=0.3\textwidth]{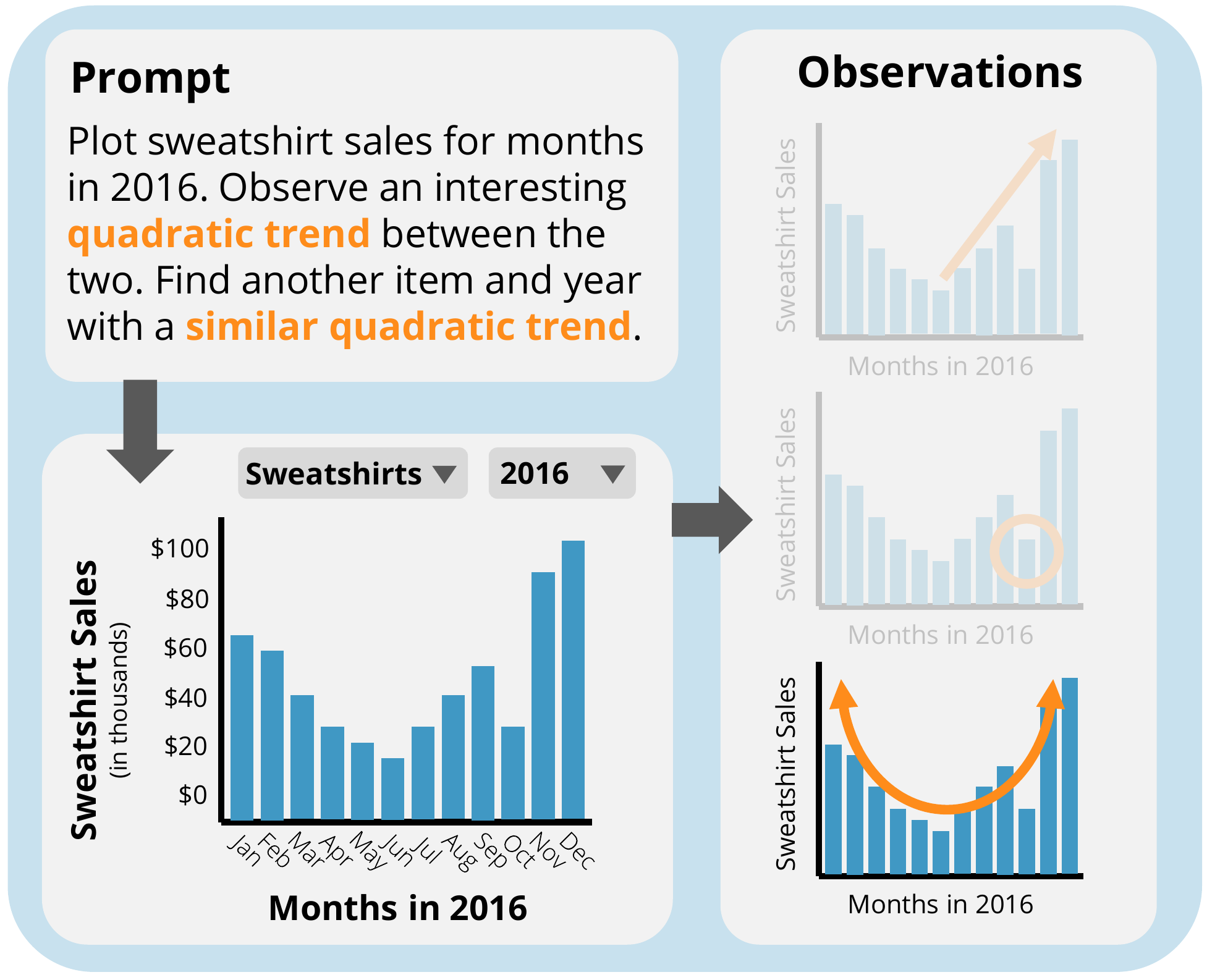} }
    \label{fig:tuning_hypothesis_space_c}
    }
    \vspace{-13px}
    \caption{\textit{Demonstration of tuning the complexity of an inferential task by modifying the specificity of the prompt. 
  (a) shows a vague prompt instructing participants to observe an interesting ``relation'' in the chart. The open-ended nature of this task results in participants reasoning through many potential characteristics in the visualization.
    (b) is a less vague prompt instructing participants to observe an interesting ``linear trend'', supplying a more specific relation to identify in the visualization.
    (c) is the most explicit prompt, instructing participants to observe an interesting ``quadratic trend.'' The specificity of this prompt gives participants the most evidence for their target of inference.}
    }
    \label{fig:tuning_hypothesis_space}
\end{figure*}

\vspace{-15px} 

\section{Formalization of Inferential Tasks}
\label{sec:itasks}
We formalize Green et al.'s and Ziemkiewicz et al.'s previous use of inferential tasks as a procedure consisting of four stages. Figure~\ref{fig:itask_teaser} illustrates this process. Although the authors do not describe how to construct inferential tasks for visualization evaluation, Green et al. do provide an example of the task set-up: an `exemplar' is first shown to participants who are then asked to ``\textit{find another example that shares/does not share a variety of characteristics.}'' In this section, we outline the four stages that make up an inferential task and define terminology to aid researchers in deploying inferential tasks in practice. We generalize inferential tasks for interactive visualization tools; however, researchers can modify specifics at each stage if the experiment goal is to evaluate one or more static visualization(s).


\subsection{Stage 1: Generate Visualization \texorpdfstring{\faBarChart}{Generate Visualization Lg}}
\label{subsec:generateVis}

\noindent
\textbf{Visual Target:} A specific visualization participants are instructed to generate (if using an interactive system) and observe.

At the start of an inferential task, participants are instructed to construct and inspect a \textit{specific visualization} -- the \textbf{visual target} for the task. For example, if using the tool in Figure~\ref{fig:inferentialTaskExample} to solve the task \iTaskExampleMidSentence participants will generate a visualization with sweatshirts selected as Item and 2016 selected as Year. The chart shown in Figure~\ref{fig:inferentialTaskExample} is the visual target for the above task. 

The purpose of having participants generate a visual target is to test the usability of a visualization tool and the design of a visual encoding. 
This stage is crucial when there are multiple visual encoding options for generating the visualization. When appropriate, researchers may specify the visual encoding that participants should generate for the visual target in the task prompt (e.g., build a pie chart for one task and a bar chart in another). If the participant is not working with an interactive system, the researcher can supply the visual target instead. In this case, the task begins at Stage 2.

\subsection{Stage 2: Enumerate Characteristics \texorpdfstring{\faCogs}{Enumerate Characteristics Lg}}

\textbf{Candidate Observation:} A proposed explanation or characterization of the data that is deduced from the visual target.


Once participants have constructed (or been given) the visual target, they are asked to find another example of a visualization that displays \textit{similar or dissimilar characteristics}. This process requires participants to identify and enumerate all plausible patterns or relations shown in the visual target that could be found elsewhere in the data. We call the particular characteristic, pattern, or relation that participants discern from the visual target their \textbf{candidate observation}. 
To illustrate how a participant arrives at their candidate observation, consider Figure~\ref{fig:tuning_hypothesis_space_a}. A participant may pose the following observations for relations in monthly sweatshirt sales:

\begin{itemize}[leftmargin=*,topsep=0pt, partopsep=0pt,itemsep=1pt,parsep=0pt]
    \item Sales spike at the end of the year (i.e., last two bars are highest)
    \item October is an outlier in the fall (i.e., one bar seems out of place)
    \item Sales decrease then increase (i.e., the bars go down then up)
\end{itemize}

Ultimately, the participant will choose a particular characterization of the visual target that they think best exemplifies a relation in monthly sweatshirt sales.
During this process, participants are implicitly performing a series of analytic tasks (e.g., ``correlate'', ``find anomalies''\cite{amar2005low}) to identify their candidate observation. 

Researchers can adjust the specificity of the task prompt to broaden or restrict the analytic tasks a user performs, and thus the number of observations a participant makes with the visual target. Figure~\ref{fig:tuning_hypothesis_space} demonstrates this mechanism. We note that an overly ambiguous task may result in HARKing\cite{kerr1998harking} (i.e., identifying a candidate observation \textit{after} completing the task) or the multiple-comparison problem in visual analysis\cite{zgraggen2018investigating}.
To lower this risk, participants can be asked to report their candidate observation before continuing with the task (e.g., \textit{experimental preregistration}\cite{cockburn2018hark}), or, the researcher could provide predefined multiple choice candidate solutions for participants to choose from.


\subsection{Stage 3: Explore Unknown Data \texorpdfstring{\faSearch}{Explore Unknown Data Lg}}
\textbf{Proposed Solution:} A visualization, showing different data attributes than the visual target, that the participant believes to exhibit the same characteristics as the candidate observation.

After identifying a candidate observation, participants explore the remaining data to find a different subset of data that, when visualized, exhibits those characteristics. As part of this process, participants generate and/or observe visualizations to reason about previously unknown data. Each of these visualizations is a \textbf{proposed solution} -- or potential answer -- to the inferential task. 

For example, take again the inferential task prompt and visualization tool shown in Figure~\ref{fig:tuning_hypothesis_space_a}. Suppose that a participant performing this task inspects the visual target and forms the observation: ``Sales spike at the end of the year.'' The next step will be to search all possible combinations of Items and Years until finding another instance for which sales spike at the end of the year.
During this search, participants will construct new visualizations with different Items and Years than those given in the prompt, i.e., proposed solutions.    

When instructing participants to ``\textit{find another example}'', researchers can specify in the prompt which data should be explored for an answer to the task. Modifying the breadth of data to search in an inferential task helps evaluate a visualization's scalability. This includes the evaluation of the visualization in helping a participant navigate through large and high-dimensional spaces, as well as testing the visualization's ability to support a participant in reasoning about relationships between many attributes at a time. 

\subsection{Stage 4: Validate Findings \texorpdfstring{\faCheck}{Validate Findings Lg}}
\textbf{Answer:} The final visualization or solution, as validated by the participant, that is believed to correctly exhibit similar characteristics as the candidate observation and visual target.

At this stage of an inferential task, the participant has a proposed solution in mind that needs to be validated for correctness as a potential \textbf{answer} to the task. The validation process requires the participant to compare their proposed solution to the original visual target. This process results in three possible outcomes, illustrated as the three outgoing arrows from Stage 4 in Figure~\ref{fig:itask_teaser}.


\begin{itemize}[leftmargin=*,topsep=0pt, partopsep=0pt,itemsep=1pt,parsep=0pt]
    \item The participant finds the proposed solution to be satisfactory. In this case, the participant has found a particular visualization that exhibits similar characteristics as the visual target.
    \item The participant is not satisfied with the current proposed solution, but believes their candidate observation is still correct. In this case, the participant will continue to explore the remaining data to find a visualization that better fits their candidate observation.
    \item The participant is not satisfied with the proposed solution, and believes that their candidate observation is incorrect. In this case, the participant will return to the second stage to identify a new candidate observation.
\end{itemize}

The process of validating a candidate observation is crucial to the evaluation of a visualization with inferential tasks, as it ensures the visualization tool (or static graphic) is capable of supporting new insights through exploratory and/or confirmatory analysis. 
We discuss the practicality of assessing participants' answers in Section~\ref{sec:discussion}.


\section{Demonstration and Validation}
\label{sec:validation}

To demonstrate how our framework for inferential tasks can be used for evaluating visualizations in practice, we present a crowdsourced study comparing the performance of two well-known and open-source visualization tools, Polestar\cite{2016-voyager} and Voyager 2\cite{2017-voyager2}, using an inferential task-based evaluation. 

Voyager 2 is a visual analytics tool that is designed to both manually and automatically support analysts through open-ended and focused exploration. 
Details are available at \URL{https://github.com/vega/voyager}. In the original user study for Voyager 2\cite{2017-voyager2}, the authors conducted a think-aloud protocol comparing against Polestar - a visualization tool similar to Tableau\cite{stolte2002polaris}. 
Unlike Voyager 2, Polestar does not provide recommendations to a user. We posit that participants will perform better on inferential tasks using Voyager 2 due to its recommendation engine.


\noindent 
\textbf{Tasks:} 
In our user study, each inferential task is structured with our formalism in mind: participants are asked to plot two specific data attributes (Generate Visualization), observe a relation in the chart (Enumerate Characteristics), and find one or two other attributes (Explore Unknown Data) that exhibit a similar relationship when visualized (Validate Findings). Participants completed their tasks with the \textit{movies} dataset using either Voyager 2 or Polestar.



\begin{itemize}[leftmargin=*,topsep=0pt, partopsep=0pt,itemsep=1pt,parsep=0pt]
    \item[] \textbf{Task 1:} Plot IMDB Votes on the x-axis and IMDB Rating on the y-axis. Observe the relationship of these variables. Find another variable that shows a similar relationship with IMDB Rating. 
    \item[] \textbf{Task 2:} Plot US Gross on the x-axis and Worldwide Gross on the y-axis. Observe the relationship of these variables. Find two different variables that show a similar relationship.
\end{itemize}

For Task 1, participants are asked to explore up to $6$ combinations of attributes on the x-axis (IMDB Rating fixed on y-axis), while Task 2 asks participants to explore up to $30$ combinations of attributes on the x- and y-axis. 
We determined ground truth by identifying data attribute(s) that display a clearly similar relationship (logarithmic for Task 1, positive linear for Task 2) to the original visualization, i.e., visual target. Task 1 had two possible correct answers, while Task 2 had only one. Examples of correct and incorrect answers are provided in the supplemental.
Accuracy was recorded as a binary `correct' or `incorrect', and speed was recorded as the total time spent between starting the task and submitting an answer. 

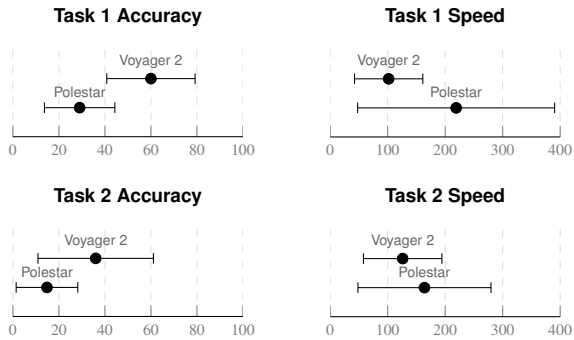
\begin{figure}[tb]
     \centering
     \begin{subfigure}[b]{.5\linewidth}
         \centering
         \usepgfplotslibrary{colorbrewer}

\pgfplotsset{compat=1.7,
    minor grid style={grey!25},
    major grid style={dashed,grey!25},
    every tick label/.append style={font=\footnotesize, grey},
    myerr/.append style={only marks,error bars/.cd, y dir=both,y explicit, x dir=both,x explicit},
    every axis/.append style={label style={font=\footnotesize}},
}

\sffamily
\begin{tikzpicture}
    \begin{axis}[
        axis lines*=left,
        axis y line=none,
        title = {Task 1 Accuracy},
        title style={font=\small\bfseries},
        xbar,
        xmajorgrids,
        width=1.1\linewidth,
        height=0.65\linewidth,
        ytick={0,1,2},
        tick label style={font=\tiny},
        xlabel style={font=\tiny},
        yticklabels={
            ,
            Voyager 2,
            Polestar,
        },
        ymin=0,
        ymax=3,
        xmin=0,
        xmax=100,
        cycle list/Paired,
        nodes near coords align={horizontal}
        ]
        \addplot[
        scatter, 
        only marks, 
        visualization depends on=\thisrow{alignment} \as \alignment,
        nodes near coords,
        point meta=explicit symbolic,
        every mark/.style={fill=black, draw=none},
        every node near coord/.style={anchor=\alignment, font=\tiny, yshift=1pt, black!60},
        error bars/.cd,
        x dir=both,
        x explicit
        ] 
        table [
        meta index=3, x error=xerror
        ] {
        x   y  xerror label    alignment 
        60  2  19.2 {Voyager 2} 270 
        29  1 15.3 {Polestar}    270 
        };
    \end{axis}
\end{tikzpicture}
\normalfont
     \end{subfigure}%
     \hfill%
     \begin{subfigure}[b]{.5\linewidth}
         \centering
         \usepgfplotslibrary{colorbrewer}

\pgfplotsset{compat=1.7,
    minor grid style={grey!25},
    major grid style={dashed,grey!25},
    every tick label/.append style={font=\footnotesize, grey},
    myerr/.append style={only marks,error bars/.cd, y dir=both,y explicit, x dir=both,x explicit},
    every axis/.append style={label style={font=\footnotesize}},
}

\sffamily
\begin{tikzpicture}
    \begin{axis}[
        axis lines*=left,
        axis y line=none,
        title = {Task 1 Speed},
        title style={font=\small\bfseries},
        xbar,
        xmajorgrids,
        width=1.1\linewidth,
        height=0.65\linewidth,
        ytick={0,1,2},
        tick label style={font=\tiny},
        xlabel style={font=\tiny},
        yticklabels={
            ,
            Voyager 2,
            Polestar,
        },
        ymin=0,
        ymax=3,
        xmin=0,
        xmax=400,
        nodes near coords align={horizontal}
        ]
        \addplot[
        scatter, 
        only marks, 
        visualization depends on=\thisrow{alignment} \as \alignment,
        nodes near coords,
        point meta=explicit symbolic,
        every node near coord/.style={anchor=\alignment, font=\tiny, yshift=1pt, black!60},
        error bars/.cd,
        x dir=both,
        x explicit
        ] 
        table [
        meta index=3, x error=xerror
        ] {
        x      y  xerror label       alignment 
        101.7  2  59.4   {Voyager 2} 270 
        219.1  1  171.3  {Polestar}  270 
        };
    \end{axis}
\end{tikzpicture}
\normalfont
     \end{subfigure}%
     
    \medskip
    
    \begin{subfigure}[b]{.5\linewidth}
         \centering
         \pgfplotsset{compat=1.7,
    minor grid style={grey!25},
    major grid style={dashed,grey!25},
    every tick label/.append style={font=\footnotesize, grey},
    myerr/.append style={only marks,error bars/.cd, y dir=both,y explicit, x dir=both,x explicit},
    every axis/.append style={label style={font=\footnotesize}},
}

\sffamily
\begin{tikzpicture}
    \begin{axis}[
        axis lines*=left,
        axis y line=none,
        title = {Task 2 Accuracy},
        title style={font=\small\bfseries},
        xbar,
        xmajorgrids,
        width=1.1\linewidth,
        height=0.65\linewidth,
        ytick={0,1,2},
        tick label style={font=\tiny},
        xlabel style={font=\tiny},
        yticklabels={
            ,
            Voyager 2,
            Polestar,
        },
        ymin=0,
        ymax=3,
        xmin=0,
        xmax=100,
        nodes near coords align={horizontal}
        ]
        \addplot[
        scatter, 
        only marks, 
        visualization depends on=\thisrow{alignment} \as \alignment,
        nodes near coords,
        draw=black,
        point meta=explicit symbolic,
        every mark/.style={fill=black, draw=none},
        every node near coord/.style={anchor=\alignment, font=\tiny, yshift=1pt, black!60},
        error bars/.cd,
        x dir=both,
        x explicit
        ] 
        table [
        meta index=3, x error=xerror
        ] {
        x     y   xerror  label        alignment 
        36    2   25.09   {Voyager 2}  270 
        14.8  1   13.4    {Polestar}   270 
        };
    \end{axis}
\end{tikzpicture} 
\normalfont
    \end{subfigure}%
    \hfill%
    \begin{subfigure}[b]{.5\linewidth}
         \centering
         \pgfplotsset{compat=1.7,
    minor grid style={grey!25},
    major grid style={dashed,grey!25},
    every tick label/.append style={font=\footnotesize, grey},
    myerr/.append style={only marks,error bars/.cd, y dir=both,y explicit, x dir=both,x explicit},
    every axis/.append style={label style={font=\footnotesize}},
}

\sffamily
\begin{tikzpicture}
    \begin{axis}[
        axis lines*=left,
        axis y line=none,
        title = {Task 2 Speed},
        title style={font=\small\bfseries},
        xbar,
        xmajorgrids,
        width=1.1\linewidth,
        height=0.65\linewidth,
        ytick={0,1,2},
        tick label style={font=\tiny},
        xlabel style={font=\tiny},
        yticklabels={
            ,
            Voyager 2,
            Polestar,
        },
        ymin=0,
        ymax=3,
        xmin=0,
        xmax=400,
        nodes near coords align={horizontal}
        ]
        \addplot[
        scatter, 
        only marks, 
        visualization depends on=\thisrow{alignment} \as \alignment,
        nodes near coords,
        draw=black,
        point meta=explicit symbolic,
        every mark/.style={fill=black, draw=none},
        every node near coord/.style={anchor=\alignment, font=\tiny, yshift=1pt, black!60},
        error bars/.cd,
        x dir=both,
        x explicit
        ] 
        table [
        meta index=3, x error=xerror
        ] {
        x      y  xerror  label       alignment 
        126.1  2  68.2   {Voyager 2}  270 
        164    1  115.7  {Polestar}   270 
        };
    \end{axis}
\end{tikzpicture} 
\normalfont
    \end{subfigure}%
    \caption{\textit{Voyager 2 vs. Polestar experiment results on both tasks. Mean accuracy and 95\% CIs are shown on the left. Mean speed (s) and standard deviation is shown on the right.
    }}
    \vspace{-5pt}
    \label{fig:voyagerVSPolestarResults}
\end{figure}

\noindent 
\textbf{Results \& Takeaways: }
Our quantitative results are summarized in Figure~\ref{fig:voyagerVSPolestarResults}, and a write-up comparing our results to those of the original study is included in the supplemental.  Overall, we find that participants are more accurate with Voyager 2 than Polestar for Task 1 and faster with Voyager 2 for both Task 1 and Task 2.  Our findings suggest that the recommendations of Voyager 2 are of high quality and assist participants in navigating through data efficiently.

Our results also highlight two potential limitations of the approach. 
First, the overall accuracy is low, particularly for Task 2. 
This suggests that asking participants to explore many combinations of attributes ($6$ in Task 1 versus $30$ in Task 2) could result in poorer accuracy. 
Second, because of the nature of our crowdsourced study, we cannot know precisely \textit{when} participants failed in their tasks. Fine-tuning the complexity of the experiment (e.g., by providing multiple choice answers or predefined candidate observations) and analyzing interaction logs could reduce this ambiguity, thereby providing additional context to the researcher. We discuss these trade-offs and avenues for future work further in Section~\ref{sec:discussion}.

\section{Discussion, Limitations, and Future Work}
\label{sec:discussion}

Although prior work highlights the benefits of using inferential tasks in evaluating visualizations\cite{green2010using, ziemkiewicz2011locus}, they are not without limitations and shortcomings. 
For example, finding the balance between an open-ended versus narrow prompt can be difficult. On one hand, an open-ended prompt (e.g., Figure~\ref{fig:tuning_hypothesis_space_a}) necessitates participants in exploring more of the data and tool being evaluated. However, as demonstrated in Section~\ref{sec:validation}, participants' accuracy can suffer when searching many possible attributes for a correct answer. An open-ended prompt also requires researchers to identify all possible characteristics that a participant could possibly (and correctly) observe. A narrow prompt (e.g., Figure~\ref{fig:tuning_hypothesis_space_c}) can be used to reduce the complexity of the task as well as limit the number of possible correct answers. Subsequently, these tasks are less indicative of the practical use a visualization tool and closer to low-level tasks\cite{amar2005low}. Balancing inferential task complexity with feasibility needs to be carefully considered and studied given an experimental goal.

As such, inferential tasks are not intended to be a replacement for open-ended \textit{nor} low-level tasks. Instead, they should be thought of as a complementary evaluation technique that can serve as a mid-point between the two. Future work is needed to better understand when an inferential task evaluation is most appropriate. In some cases, such as a strict usability study, low-level tasks are sufficient. When interested in testing how well a visualization supports new insights, the point at which an inferential task evaluation becomes as useful as an open-ended evaluation can be unclear.  We leave to future work investigating the precise benefits and trade-offs of inferential tasks as an evaluation technique for visualization.

\section{Conclusion}
\label{sec:conclusion}
In conclusion, we formalize the use of inferential tasks as a way to build on complex benchmark tasks -- creating evaluations that can be quantitatively analyzed, yet engage participants in a pattern of analysis closer to open-ended tasks for insight-based evaluations. 
We present a crowdsourced study to demonstrate the use of inferential tasks in practice with two interactive visualization tools. Our results suggest that our framework for inferential tasks can be successfully deployed to illuminate differences between visualization systems.

\vspace{-4pt}

\section*{Acknowledgments} 
This work was supported by National Science Foundation grants IIS1452977, OAC-1940175, OAC-1939945, DGE-1855886, OAC-2118201, NRT-2021874, and DOD grants
HQ0860-20-C-7137, N68335-17-C-0656. We thank the reviewers for their feedback.

\bibliographystyle{eg-alpha-doi}
\bibliography{itasks}


\end{document}